\documentclass[12pt]{article}
\usepackage{graphicx}

\newcommand\pubnumber{SNSN-323-63}
\newcommand\pubdate{\today}

\def\lbnl{Nuclear Science Division\\
Lawrence Berkeley National Laboratory, Berkeley, CA 94720, USA}

\def\Title#1{\begin{center} {\Large #1 } \end{center}}
\def\Author#1{\begin{center}{ \sc #1} \end{center}}
\def\Address#1{\begin{center}{ \it #1} \end{center}}

\newcommand\pubblock{\rightline{\begin{tabular}{l} \pubnumber\\
         \pubdate  \end{tabular}}}
\newenvironment{Abstract}{\begin{quotation}  }{\end{quotation}}
\newenvironment{Presented}{\begin{quotation} \begin{center} 
             PRESENTED AT\end{center}\bigskip 
      \begin{center}\begin{large}}{\end{large}\end{center} \end{quotation}}

\begin{document}
\begin{titlepage}
\pubblock

\vfill
\Title{Recent Results on Heavy Quark Production in High Energy Nucleus Collisions}
\vfill
\Author{Xin Dong}
\Address{\lbnl}
\vfill
\begin{Abstract}
The goal of the ultra-relativistic heavy ion program is to study Quantum Chromodynamics under finite temperature and density conditions. After a couple of decades of experiment, the focus at the top RHIC and the LHC energy has evolved to quantitative understanding of properties of the hot and dense medium, namely the strongly-coupled Quark Gluon Plasma (sQGP) created in these heavy-ion collisions, and to constrain transport parameters of the sQGP medium. Heavy quarks offer unique insights towards detailed understanding of the sQGP properties due to their large masses. Recent collider and detector advances have enabled precision measurements of heavy quark hadron production in heavy-ion collisions.

In this talk, I will review recent results from heavy quark production measurements at RHIC and the LHC. These high quality data will offer stringent constraints on theoretical model calculations and help precision determination of QGP medium transport parameters. Finally, I look forward to a more prospective future of the heavy quark program with further improved detectors at both RHIC and the LHC.
\end{Abstract}
\vfill
\begin{Presented}
Thirteenth Conference on the Intersections of Particle and Nuclear Physics (CIPANP 2018)\\
\vskip 0.1in
Palm Springs, CA, USA,  May 29 -- June 3, 2018
\end{Presented}
\vfill
\end{titlepage}
\def\thefootnote{\fnsymbol{footnote}}
\setcounter{footnote}{0}

\section{Introduction}
\label{sec:intro}
The ultra-relativistic heavy-ion collisions (uRHICs) provide a unique opportunity to study Quantum Chromodynamics (QCD), the theory that describes strong interactions between quarks and gluons, under the extreme temperature and density regions in a terrestrial laboratory. Lattice QCD (lQCD) predicts that under such conditions nuclear matter will undergo a phase transition into a plasma of quarks and gluons (QGP). Recent calculations 
with the full QCD Lagrangian with (2+1) flavors and physical quark masses show the presence of a cross-over phase transition from hadronic matter to the QGP phase at a critical temperature $T_{c}$ = 154 $\pm$ 9\,MeV ($\sim$1.8$\times10^{12}$\,K) in the region of zero baryon density~\cite{Bazavov:2014pvz,Ding:2015ona}. Such a form of matter should therefore have permeated the early universe
in the first several microseconds of its expansion~\cite{Castorina:2015ava}.

To characterize particle production in heavy-ion collisions, widely used observables are the nuclear modification factor $R_{\rm AA}$ defined as
\[
R_{\rm AA} = \frac{1}{T_{\rm AA}}\frac{d^2N^{\rm AA}dp_{\rm T}dy}{d^2\sigma^{pp} dp_{\rm T}dy}
\]
the ratio between the production yield in heavy-ion A+A collisions divided by the cross section in $p+p$ collisions normalized by the nuclei thickness function $T_{\rm AA}$, and the anisotropic flow parameters $v_n$ which are defined as Fourier coefficients of final state particle azimuthal angle distributions with respect to collision event plane.
\[
\frac{d^3N}{p_{\rm T}dp_{\rm T}dyd\phi} = \frac{d^2N}{2\pi p_{\rm T}dp_{\rm T}dy} \bigg[ 1 + \sum_{n=1}^{\infty} 2v_n\cos\big(n(\phi-\Psi_{\rm EP})\big) \bigg]
\]
The $n$-th harmonic flow coefficient $v_n$ can be calculated as $v_n = \langle\cos(n(\phi-\Psi_{\rm EP}))\rangle$ while $v_2$ conventionally called elliptic flow is often the largest coefficient in non-central heavy-ion collisions.

Over the past couple of decades, experimental results from Relativistic Heavy Ion Collider (RHIC) and the Large Hadron Collider (LHC) have shown distinguished evidences that support the formation of the new QGP phase~\cite{StarWhitePaper,PhenixWhitePaper,LhcSummary}. Out of these evidences, two key observations are
\begin{itemize}
\item {\bf Jet Quenching} - $R_{\rm AA}$ of light and strange flavor hadrons show a strong suppression at $p_{\rm T}$$>$$\sim$6--8\,GeV/$c$ in central heavy-ion collisions indicating the media created in these collision are very dense and opaque to energetic partons.
\item {\bf Partonic Collectivity} - Light and strange hadrons including $\phi$, $\Omega$ show strong elliptic flow $v_2$ and their $v_2$ follow a Number-of-Constituent-Quark (NCQ) scaling suggesting the anisotropic collectivity has been built up in the parton phase.
\end{itemize}

The physics goal of high energy heavy-ion collisions has evolved to quantitatively study the inner working of the hot QGP medium and to characterize its emergent transport properties. The most interesting ones out of these are shear-viscosity-to-entropy-density ratio $\eta/s$, mean gain in transverse momentum squared per unit path length $\hat{q}$ and heavy quark spacial diffusion coefficient $2\pi TD_s$ etc. Lattice calculations for these QGP transport parameters remain a bit challenging currently though there have been great progresses in the last several years~\cite{Ding:2015ona,latticeBanerjee,latticeDing,Plumari:2011mk,Banerjee:2011ra}. It is strongly desired to pursue precision experimental measurements to compare to first principle lQCD calculations as well as to offer constraints to phenomenological models to better understand how these intrinsic transport parameters under the extreme temperature condition emerge in the QCD.

\begin{figure}[htbp] 
\begin{center}
\includegraphics[width=0.6\textwidth]{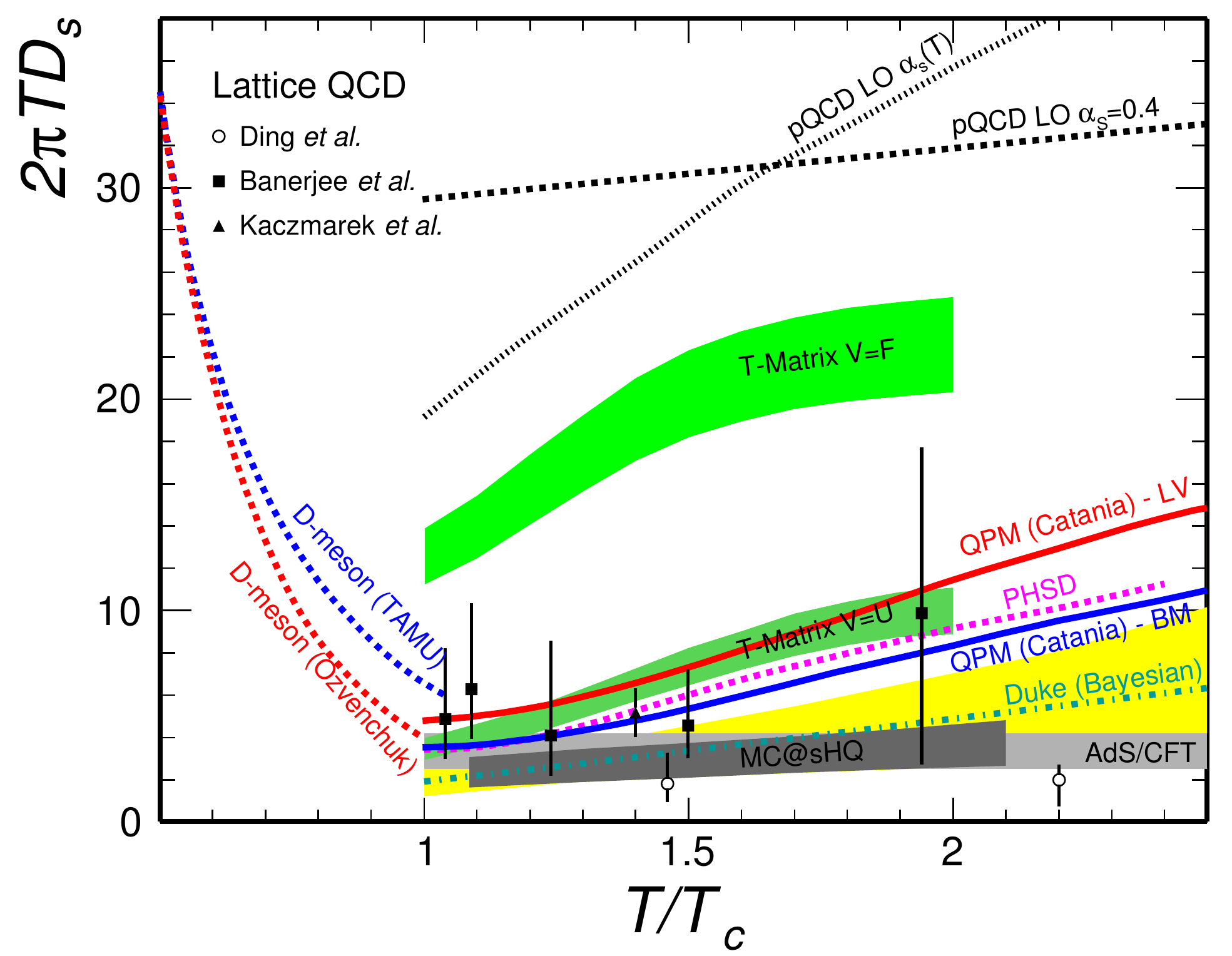}
\caption{(Color line) - Temperature dependence of charm-quark diffusion coefficients from quenched lQCD calculations~\cite{Banerjee:2011ra,Kaczmarek:2014jga,latticeDing} and various model calculations based on different
elastic interactions in the QGP~\cite{Moore:2004tg,vanHees:2004gq,Das:2015ana,Song:2015sfa,He:2014cla,Andronic:2015wma}, and AdS/CFT estimation~\cite{Horowitz:2015dta}. The yellow band shows a Bayesian analysis fit result (90\% C.R.) based on the Duke dynamic model~\cite{Xu:2017obm}. Also shown on the figure are the $D_s$ coefficients for $D$-mesons in hadronic matter (color dashed lines)~\cite{He:2011yi,Tolos:2013kva}.
 }
\label{fig:Ds}
\end{center}
\end{figure}

Heavy quarks - HQs ($c$, $b$), due to their large masses, are expected to have unique roles for studying QCD in the hot QGP medium. Heavy quark interaction with the hot QGP matter should shed light on the roles of radiative energy loss vs. elastic collisional energy loss in such a medium. In particular, one should expect the mass hierarchy for the parton energy loss in QCD medium: $\Delta E_b<\Delta E_c<\Delta E_q<\Delta E_g$, and such a difference in $R_{\rm AA}$ will disappear at very high $p_{\rm T}$ as the mass effect becomes less important or negligible. HQ propagation inside the QGP medium can be treated as ``Brownian" motion when the HQ mass is much larger than the medium temperature as well as the interaction strength. Under this assumption, the HQ equation of motion can be described reliably by a stochastic Langevin simulation and characterized by one intrinsic medium transport parameter - the HQ spacial diffusion coefficient, $2\pi TD_s$. Here low $p_{\rm T}$ measurements will be more relevant for the determination of this transport parameter. 

Figure~\ref{fig:Ds} summarizes recent calculations of temperature dependence of charm-quark diffusion coefficients $2\pi TD_s$ from perturbative QCD (pQCD)~\cite{Moore:2004tg,vanHees:2004gq}, various phenomenological models~\cite{Das:2015ana,Song:2015sfa,He:2014cla,Andronic:2015wma}, an AdS/CFT estimation~\cite{Horowitz:2015dta} and a Bayesian analysis result by fitting to experimental data based on the Duke dynamic model~\cite{Xu:2017obm}. Also shown on the plots are data points from quenched lQCD calculations~\cite{Banerjee:2011ra,Kaczmarek:2014jga,latticeDing} as well as those for $D$-mesons in hadronic matter~\cite{He:2011yi,Tolos:2013kva}.

\section{Recent Achievements}
\label{sec:achievements}

Heavy quark hadrons decay shortly after they are produced ($c\tau \sim 60 - 500 \mu m$). In the early days of measurements at ISR and RHIC energies, the measurements were mostly carried out through measuring their decay lepton daughters. The decay daughters contain contributions from various charm and bottom hadron decays. Precision measurements of HQ production in high multiplicity heavy-ion collisions require separation of their decay vertices from primary collisions in order to reduce combinatorial background. Therefore, silicon pixel detectors are critical in order to achieve necessary pointing resolution in a wide kinematic region. 
Recently, a high resolution pixel detector based on the Monolithic Active Pixel Sensor (MAPS) technology was firstly applied to the STAR experiment. Its unique features - ultimate pitch size (21$\times$21 $\mu m^2$), thin material thickness (0.4\%$X_0$ per layer) are perfect fit for precision HQ measurements over a broad momentum range in heavy-ion collisions~\cite{Contin:2017mck}. 

\begin{figure}[htbp]
\center{
\includegraphics[width=0.4\columnwidth]{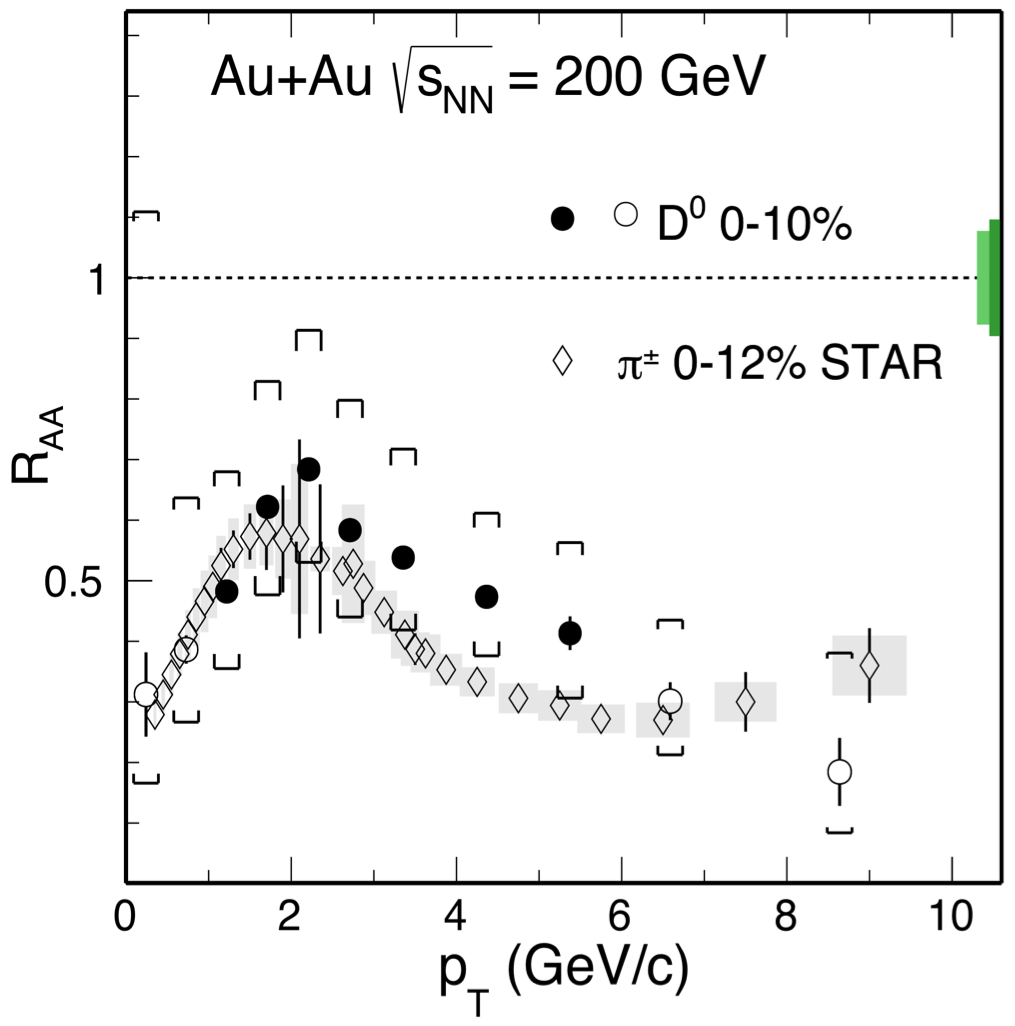}%
\hspace{0.2in}
\includegraphics[width=0.44\columnwidth]{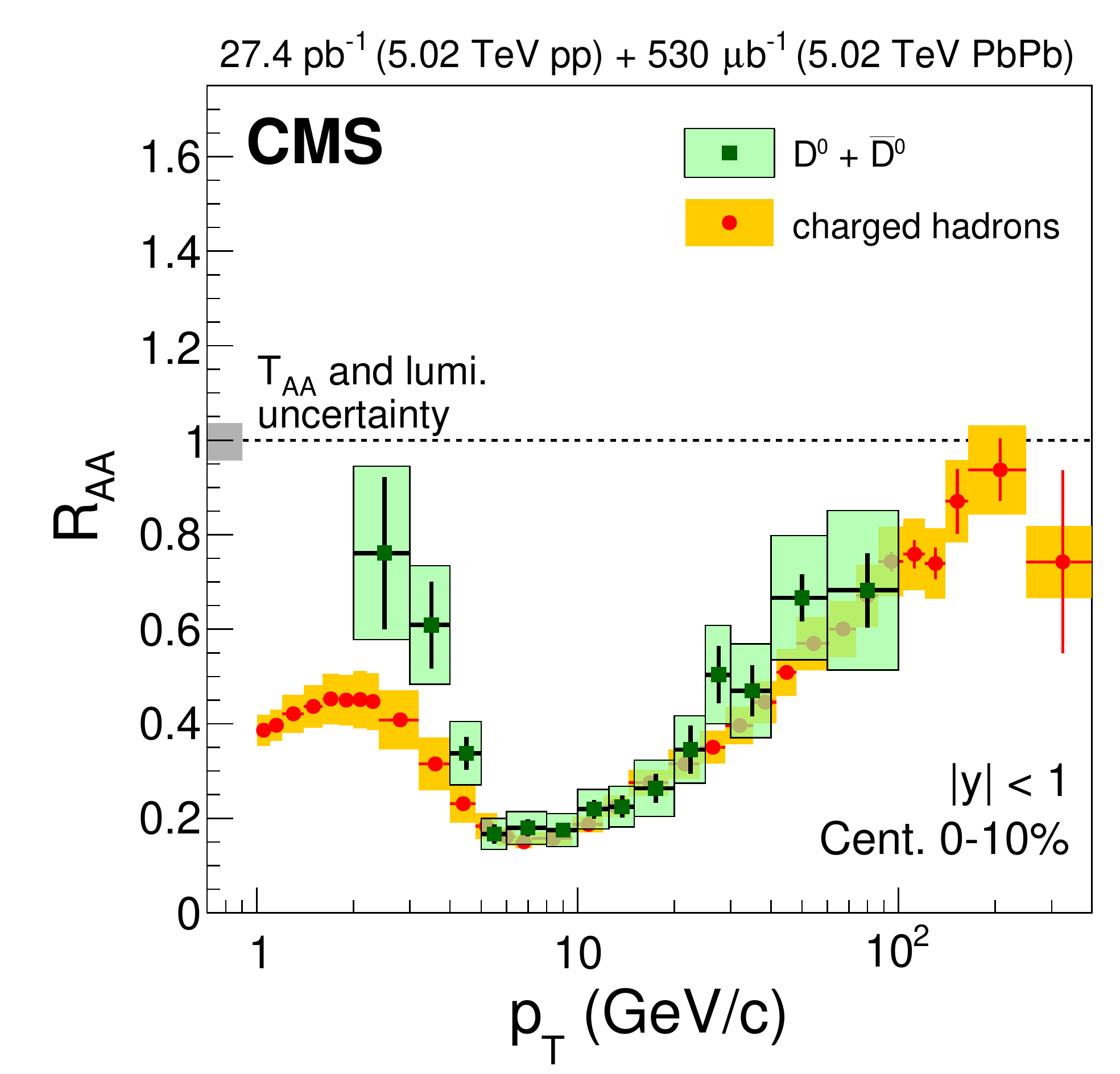}%
}
\caption{Nuclear modification factor $R_{\rm AA}$ of $D^0$ mesons in 0--10\% central heavy-ion collisions compared to charged pions at RHIC (Left) and charged hadrons at LHC (Right).
\label{fig:D0RAA}}
\end{figure}

The advanced technology enables precision measurements of various heavy flavor observables in heavy-ion collisions at both RHIC and the LHC. Figure~\ref{fig:D0RAA} left plot shows the STAR measurement of $D^0$ meson $R_{\rm AA}$ in comparison with that of charged pions in 0--10\% Au+Au collisions at $\sqrt{s_{_{\rm NN}}}$ = 200\,GeV. One can observe that $D^0$-meson $R_{\rm AA}$ is significantly suppressed at $p_{\rm T}>$ 4\,GeV/$c$ in central Au+Au collisions and the magnitude is similar as that of light hadrons. The suppression level is reduced when moving towards mid-central and peripheral collisions, consistent with the centrality dependence of light hadron suppression as well. On the other hand the $R_{\rm AA}$ of $D^0$ meson at $p_{\rm T}<$ 4\,GeV/$c$ shows an interesting bump structure in thesel Au+Au collisions. In theory model calculations, such a "flow-bump'' feature is expected from finite charm quark radial flow together with coalescence hadronization mechanism.

ALICE and CMS experiments have been leading the efforts to carry out the open HQ measurements at the LHC. Figure~\ref{fig:D0RAA} right plot shows the $R_{\rm AA}$ of inclusive $D^0$ mesons in comparison to that of charged hadrons in 0--10\% central Pb+Pb at $\sqrt{s_{_{\rm NN}}}$ = 5.02\,TeV collisions by CMS~\cite{Sirunyan:2017xss}. At $p_{\rm T}>$ 4\,GeV/$c$, one can observe that $D$-meson production in central heavy-ion collisions is significantly suppressed, and the suppression level is nearly similar compared to that of light flavor hadrons at both RHIC and LHC. Combined with no suppression seen in $p$+Pb collisions, the large suppression observed in A+A collisions demonstrates that charm quarks lose significant amount of energy in the hot QGP medium.

\begin{figure}[htbp]
\center{
\includegraphics[width=0.48\columnwidth]{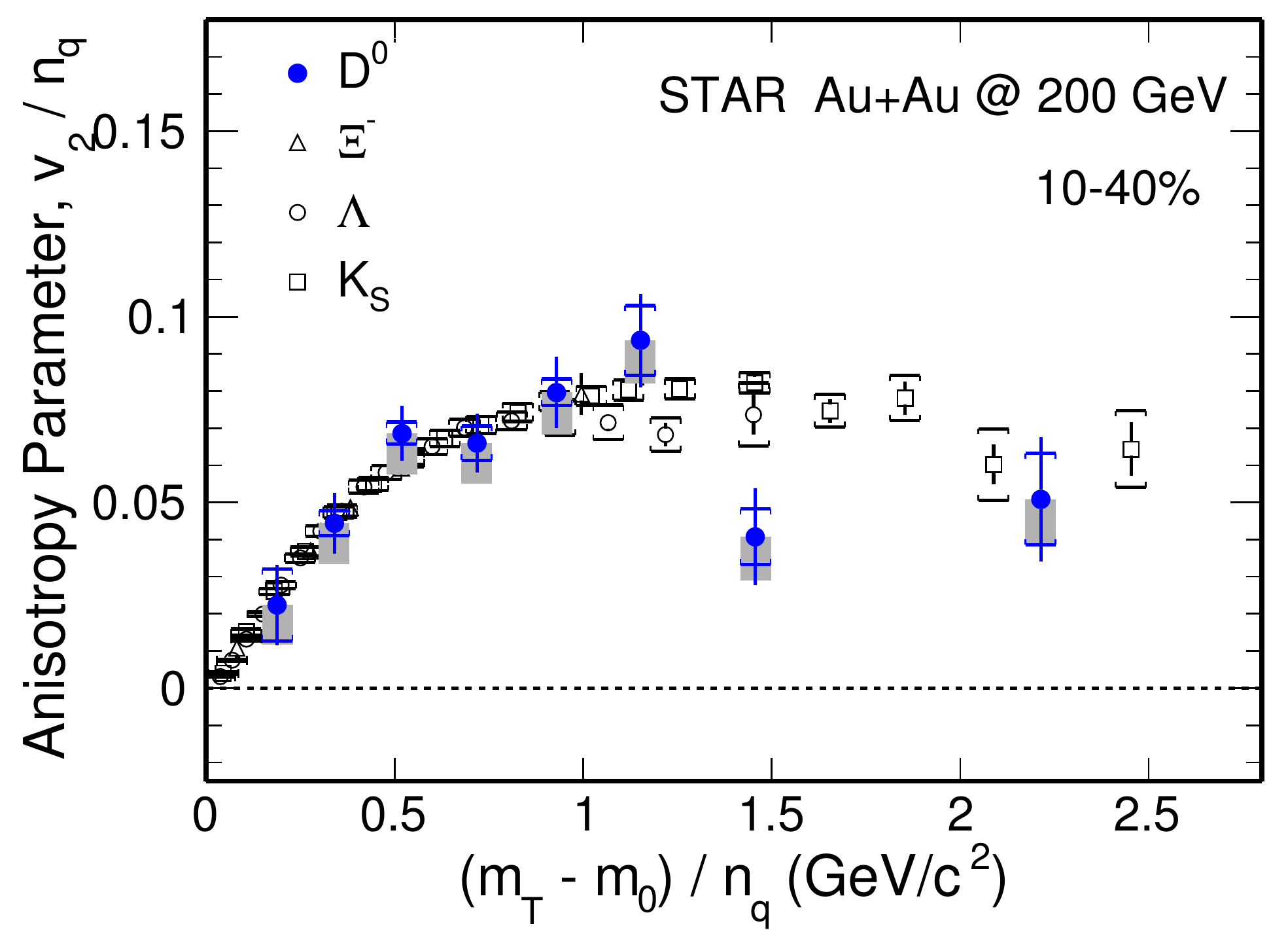}%
\includegraphics[width=0.48\columnwidth]{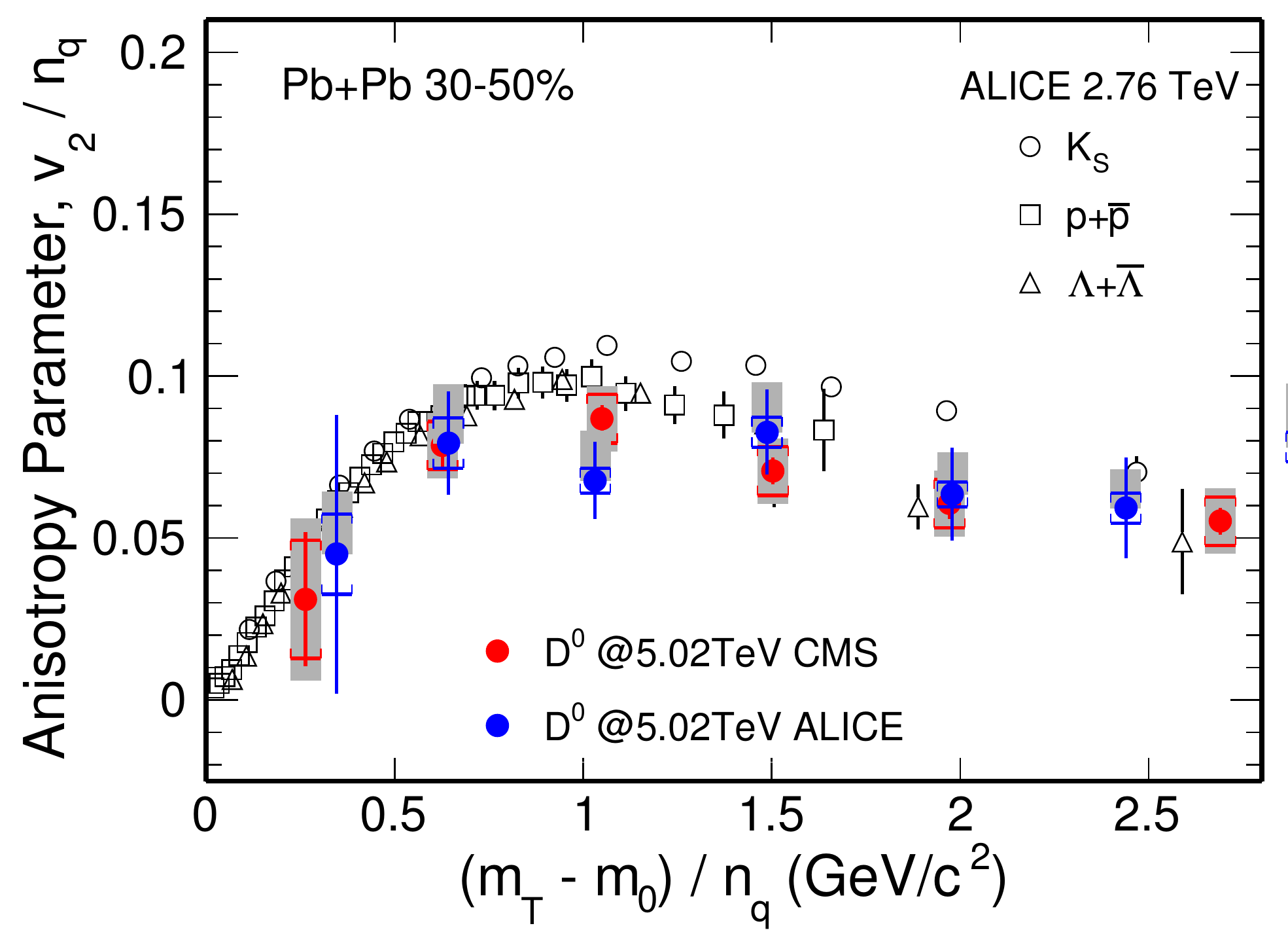}%
}
\caption{$D^0$ meson $v_2$ measured at RHIC (Left) and LHC (Right) compared to light flavor hadrons measured in mid-central heavy ion collisions. The measured $v_2$ data are scaled by the Number-of-Constituent-Quarks (NCQ, $n_q$) and plotted as a function of $n_q$ scaled transverse kinetic energy $m_{\rm T}-m_0$ while $m_{\rm T} = \sqrt{p_{\rm T}^2+m_0^2}$.
\label{fig:D0v2}}
\end{figure}

Figure~\ref{fig:D0v2} left plot shows the latest $D^0$ meson $v_2$ measurement compared to other identified particles ($K^0_{S}$, $\Lambda$ and $\Xi^-$) in 10--40$\%$ centrality Au+Au collisions at $\sqrt{s_{_{\rm NN}}}$ = 200\,GeV~\cite{Adamczyk:2017xur,Scardina:2017ipo}. The data are plotted as a function of transverse kinetic energy $m_{\rm T}-m_{0}$, where $m_{\rm T} = \sqrt{p_{\rm T}^2+m_0^2}$ and both axes are normalized by the number-of-constituent-quarks ($n_q$). One can observe that the $D^0$ mesons $v_2$ is significant and follows the same trend as light flavor hadrons. This suggests that charm quarks should have gain sufficient interactions in the QGP medium to reach such a significant flow. The right plot of Fig.~\ref{fig:D0v2} shows the $D$-meson $v_2$ measured at the LHC and compared to other identified particles as a function of $(m_{\rm T}-m_0)/n_q$ in 30--50\% centrality Pb+Pb collisions. The result demonstrates significant charm hadron $v_2$ in Pb+Pb collisions at the LHC and the magnitude is comparable to that of charged particles.  One can observe the similar universal trend for all particles measured here including $D$-mesons. There might be a small deviation from a universal trend at $1<(m_{\rm T}-m_0)/n_q<2$\,GeV/$c^2$ for the $D$-meson data at LHC, but the deviation shows also in light flavor hadrons which is argued to be due to longer hadronic stage rescatterings at the LHC.

\begin{figure}[htbp]
\center{
\includegraphics[width=0.48\columnwidth]{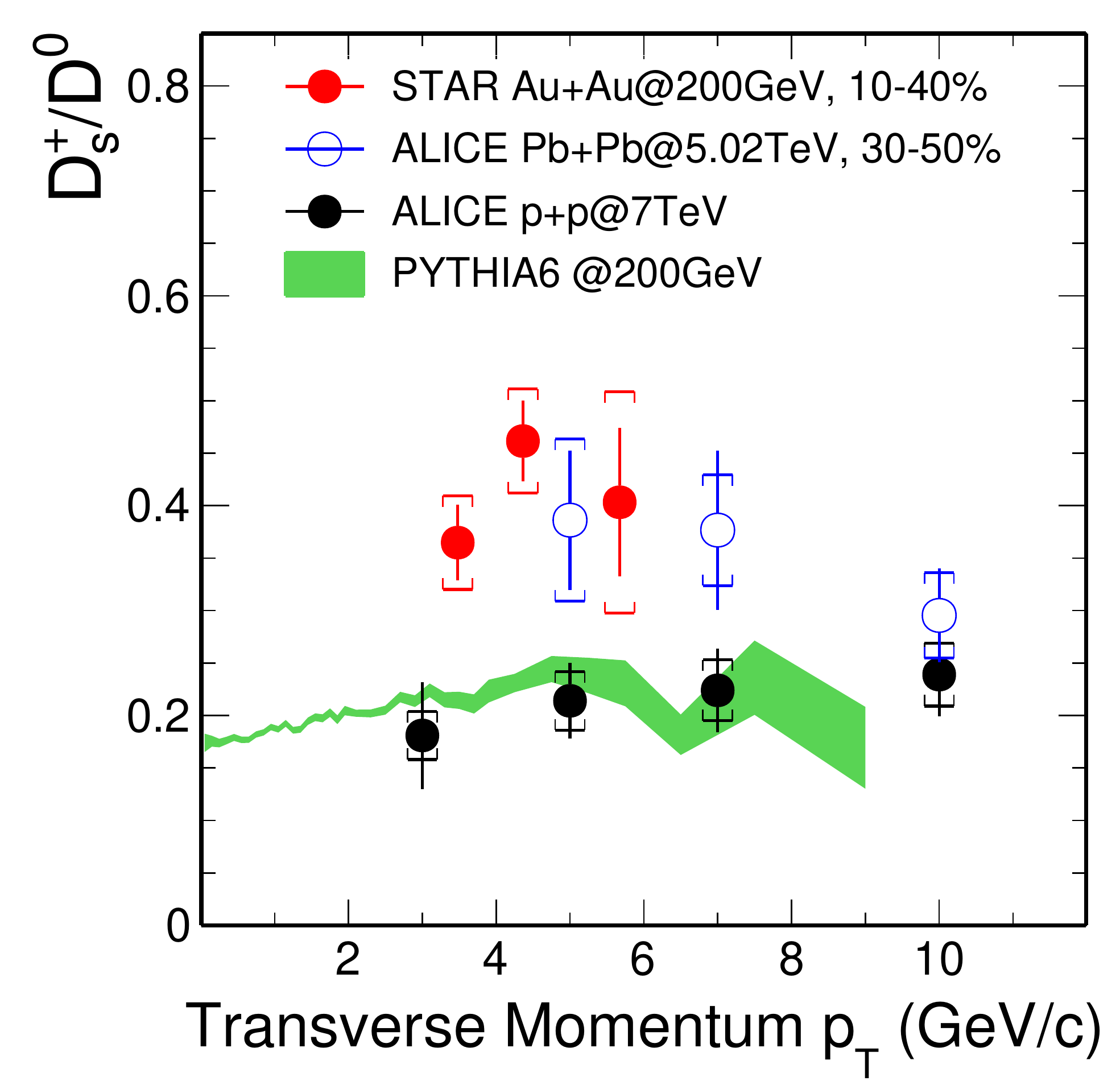}%
\includegraphics[width=0.48\columnwidth]{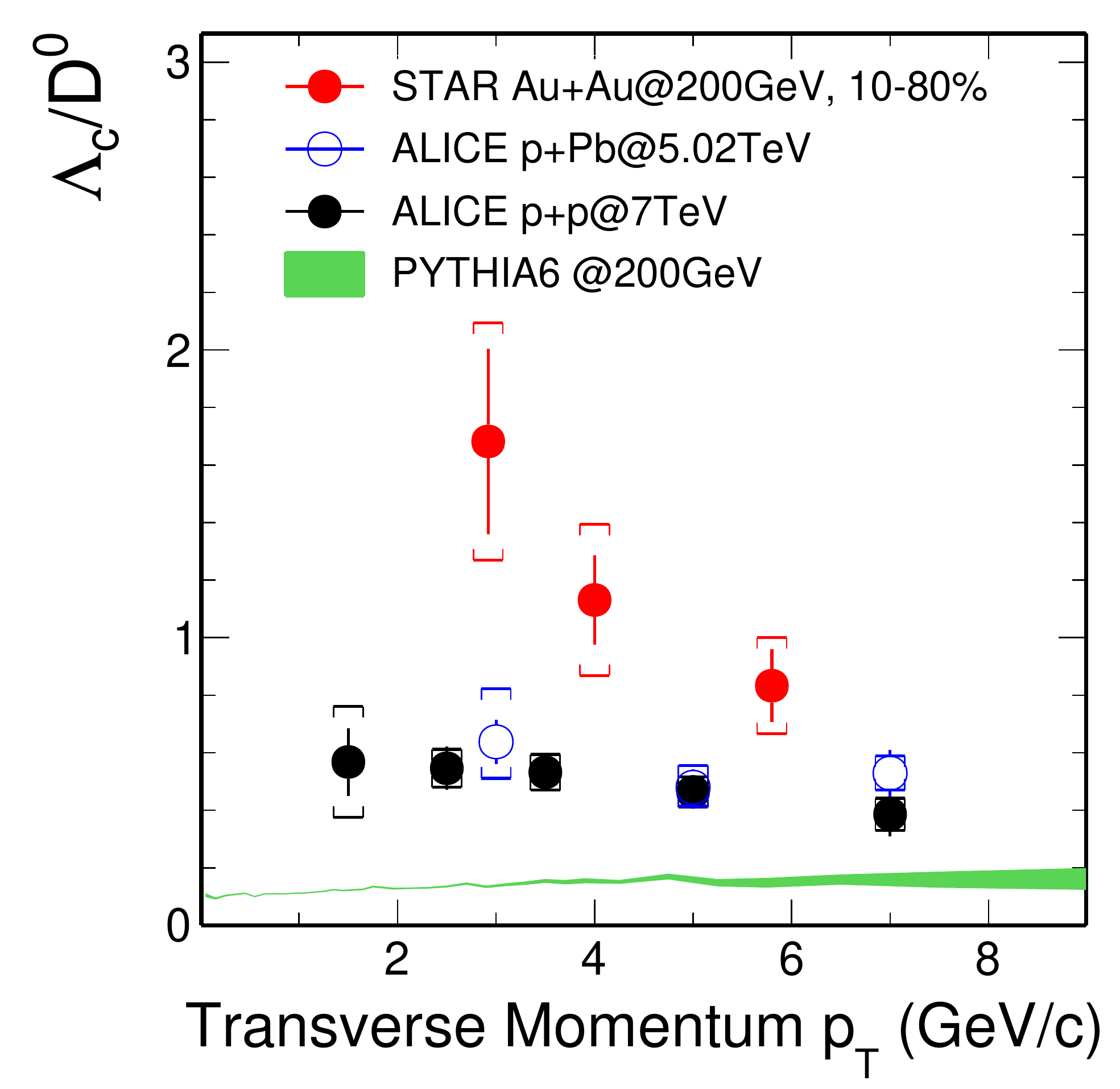}%
}
\caption{$D_s^+/D0$ (left) and $\Lambda_c^+/D^0$ ratios as a function of $p_{\rm T}$ measured in $p+p$ and heavy ion collisions by ALICE and STAR. Grey bands depict the PYTHIA6 calculations in $p+p$ collisions at $\sqrt{s_{_{\rm NN}}}$ = 200\,GeV.
\label{fig:DsLc}}
\end{figure}

In the low to intermediate $p_{\rm T}$, hadronization plays an important role in the observed charmed hadron production. Coalescence hadronization with an enhancement of strangeness production in heavy-ion collisions can lead to an enhancement of $D_s^+$ production relative to that of $D^0$-mesons in high energy heavy-ion collisions~\cite{TAMU2}. Coalescence hadronization can also lead to an enhancement of $\Lambda_c^+$ baryon production relative to other charmed mesons in central heavy-ion collisions too~\cite{Oh:2009zj,Ghosh:2014oia,Plumari:2017ntm}.

Figure~\ref{fig:DsLc} left plot shows the recent measurements of $D^+_{s}/D^0$ ratios as a function of $p_{\rm T}$ measured by STAR~\cite{STARDsLc} and ALICE~\cite{ALICEDQM17} in mid-central heavy-ion collisions compared to $p+p$ references. The PYTHIA value of $D^+_{s}/D^0$ in $p+p$ collisions is constrained using the fragmentation ratio measured in other $ep$ or $e^+e^-$ collisions, which is about 0.15-0.2. The ALICE measurement in $p+p$ collisions at $\sqrt{s}$ = 7\,TeV is consistent with the PYTHIA calculation within uncertainties. In mid-central heavy-ion collisions, the $D^+_{s}/D^0$ ratios are significantly higher than the $p+p$ reference in the measured $p_{\rm T}$ region. Such an enhancement is qualitatively consistent with the expectation from the coalescence hadronization coupled with strangeness enhancement in heavy-ion collisions. But quantitatively, the predicted $D_s^+/D^0$ ratios from model calculations~\cite{Plumari:2017ntm,He:2012df} seem to drop faster in $p_{\rm T}$ while the enhancement seen in the data still persists to the high $p_{\rm T}$.

$\Lambda_c^+$ baryon has an extreme short life time and has never been reconstructed in heavy-ion collisions. Figure~\ref{fig:DsLc} right plot shows the measured $\Lambda_c^+/D^0$ ratios at mid-rapidity in $p+p$ and $p$+Pb collisions by ALICE~\cite{Acharya:2017kfy} and in Au+Au collisions by STAR~\cite{}. The ratio measured in mid-central Au+Au collisions is significantly larger than the PYTHIA expectation or the world average fragmentation ratio from elementary collisions. It is also larger than simple statistical hadronization model estimation~\cite{Oh:2009zj}. Model calculations with coalescence hadronization for charm quarks~\cite{Oh:2009zj,Plumari:2017ntm,Minissale:2015zwa} show a better agreement with the measured data points by STAR. One also noticed that the $\Lambda_c^+/D^0$ ratio in $p+p$ and $p$+Pb collisions by ALICE is about 0.4-0.6 between 1$<p_{\rm T}<\sim$10\,GeV/$c$, much larger than the PYTHIA prediction and the average charm quark fragmentation ratio from previous experiments. PYTHIA calculations with color reconnection mechanism can lead to an increase in the $\Lambda_c^+/D^0$ ratio. However, it cannot quantitatively reproduce the measured data by ALICE.

\begin{figure}[htbp]
\center{
\includegraphics[width=0.50\columnwidth]{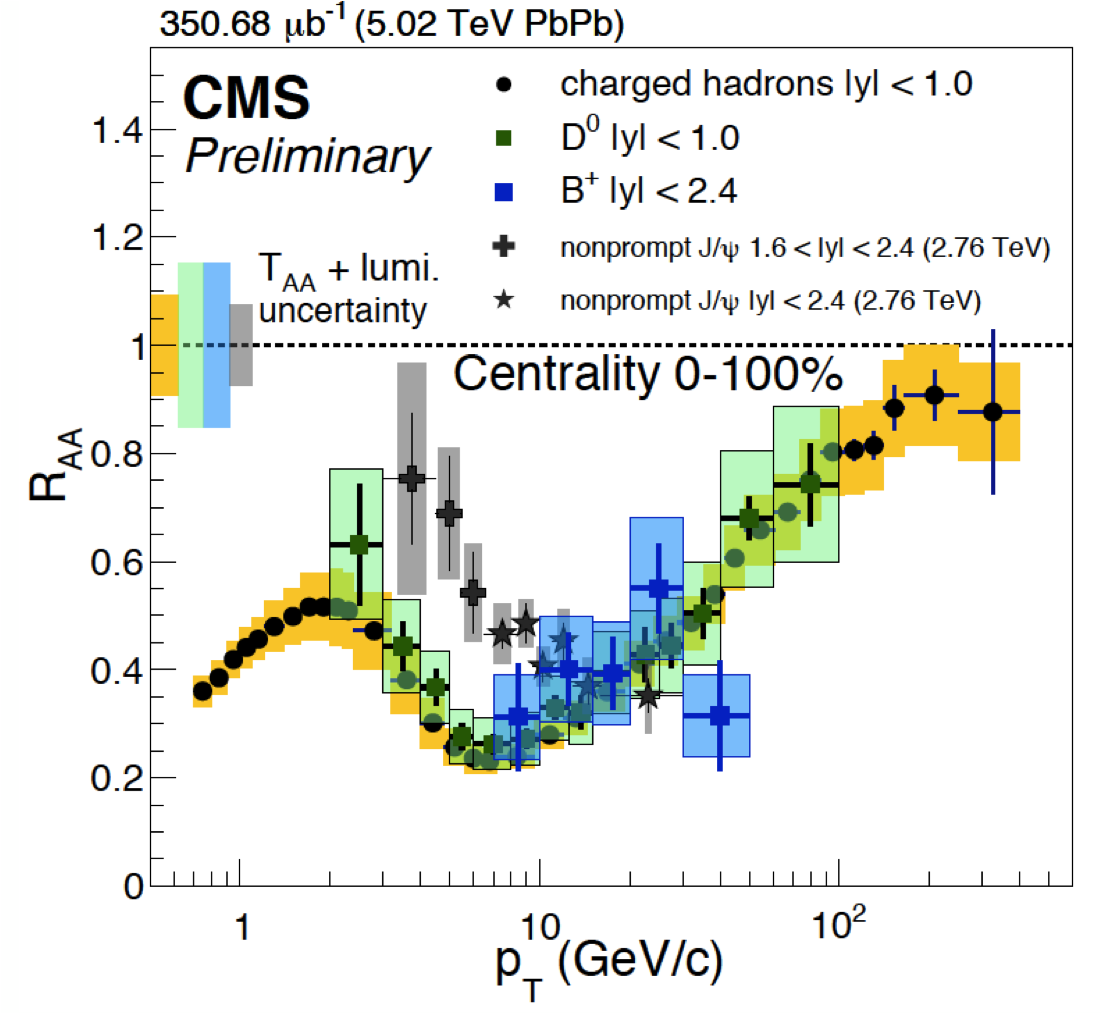}%
\includegraphics[width=0.46\columnwidth]{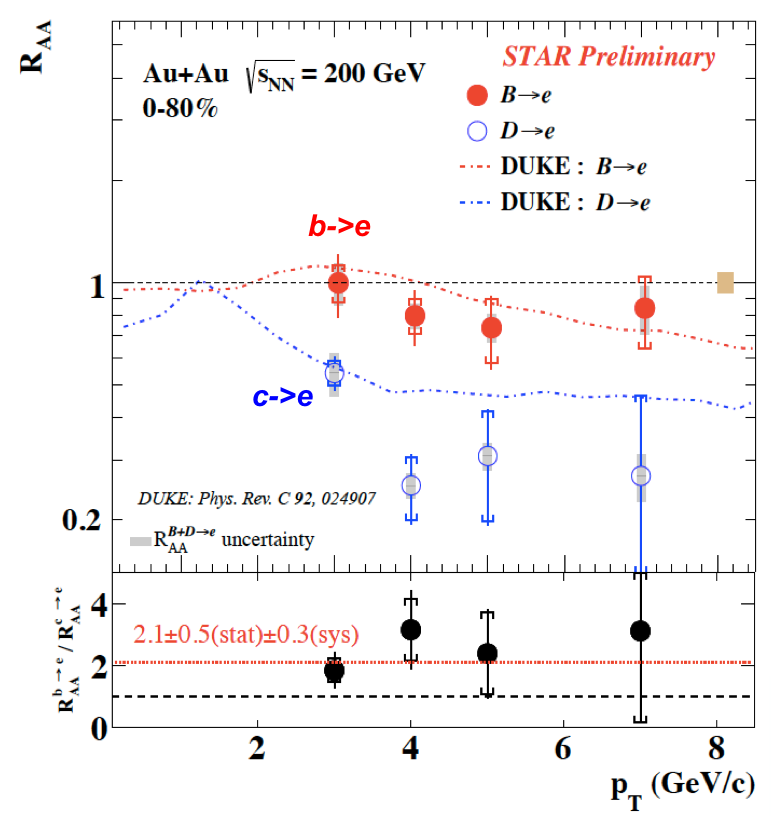}%
}
\caption{(Left) $R_{\rm AA}$ of non-prompt $J/\psi$ and fully reconstructed $B^+$-meson in Pb+Pb collisions compared to inclusive charged hadrons and $D^0$-mesons. (Right) $R_{\rm AA}$ of non-prompt $J/\psi$, non-prompt $D^0$ and charm/bottom separated electrons in Au+Au collisions by STAR.
\label{fig:BRAA}}
\end{figure}

Figure~\ref{fig:BRAA} left panel shows the $R_{\rm AA}$ of non-prompt $J/\psi$ and fully reconstructed $B^+$-meson in Pb+Pb collisions compared to inclusive charged hadrons and $D^0$-mesons in 0-100\% centrality Pb+Pb collisions at $\sqrt{s_{_{\rm NN}}}$ = 5.02\,TeV from CMS~\cite{Sirunyan:2017oug,Khachatryan:2016ypw}. In the $p_{\rm T}$ region between 8--40\,GeV/$c$, within uncertainties, the $B^+$-meson $R_{\rm AA}$ is also similar to the $R_{\rm AA}$ of charged hadrons and $D^0$ mesons. In low $p_{\rm T}$ region, the CMS collaboration measured the non-prompt $J/\psi$ $R_{\rm AA}$ and the result shows less suppressions with respect to the light hadron and $D^0$-meson $R_{\rm AA}$ in the $p_{\rm T}$ region of 4--10\,GeV/$c$ in Pb+Pb collisions at $\sqrt{s_{_{\rm NN}}}$ = 5.02\,TeV.  The figure clearly demonstrates the difference in $R_{\rm AA}$ between $B$ mesons and $D$ mesons in this kinematic region, suggesting the mass hierarchy of parton energy loss. At RHIC, both PHENIX and STAR have reported the latest measurements on the bottom production in the Quark Matter 2017 conference~\cite{PHENIXeQM17,STARBQM17}. Figure~\ref{fig:BRAA} right plot shows the STAR measurement of bottom production via charm/bottom separated electrons in Au+Au collisions. The bottom-electron $R_{\rm AA}$ is higher than the charm-electron $R_{\rm AA}$ with about a 2$\sigma$ significance.

Observations at RHIC and LHC have shown evidences that bottom $R_{\rm AA}$ is larger than those of charm and light flavor hadrons in $p_{\rm T}$$<$10\,GeV/$c$ in heavy-ion collisions, indicating bottom quarks lose less energy in the medium created in these collisions. At the LHC, the difference in $R_{\rm AA}$ between bottom hadrons or $b$-jets and charm/light flavor hadrons/jets disappears at high $p_{\rm T}$. These results are consistent with the mass hierarchy feature for parton energy loss inside the QGP medium. A detailed investigation on open bottom production in heavy-ion collisions will be necessary to evaluate quantitatively the roles between radiative energy loss vs. collisional energy loss, and to better determine the heavy quark diffusion coefficient.

\section{Open questions and future directions}
\label{sec:questions}

There have been various model calculations on the $D^0$ meson $R_{\rm AA}$ and $v_2$. However, there is still a challenge for nearly all models to simultaneously describe the $R_{\rm AA}$ and $v_2$ data, and at both RHIC and the LHC. There are quite some ingredients that may affect the charm hadron production in heavy-ion collisions. To have a coherent understanding of the heavy quark diffusion coefficient,  there are quite several important key questions related to HQ probes that need to be addressed.

\begin{itemize}
\item{{\bf Energy loss mechanisms: radiative vs. collisional?} 

Theoretically there is general consensus that up to $p_{\rm T} < 5 M_{Q}$~\cite{Cao:2016gvr} the collisional energy loss should dominate, while at higher momenta 
there is a consensus that radiative energy loss becomes dominant even if self-consistently collisional energy loss can never be discarded. 
The question then becomes how and where the radiative energy loss mechanism starts to play a significant role in the parton energy loss. The $R_{\rm AA}(p_{\rm T})$ does not appear to provide a solid signature of the
heavy quark energy loss mechanism.  Hence detailed differential studies with centrality, parton mass/flavor and even correlations could help to disentangle the two mechanisms.
}
\item{{\bf Heavy quark in-medium evolution: Langevin vs. Boltzmann?}

The original idea of using the ``Brownian motion" analogy to emulate heavy quark in-medium evolution is based on that the heavy quark mass is much larger than the medium temperature as well as the interaction strength between heavy quark and medium. Therefore the Boltzmann transport equation can be reduced to a Langevin equation and one can perform stochastic Monte Carlo simulations to reliably trace the evaluation of heavy quark propagating inside the QGP medium. Model calculations show that the difference in the charm quark spectrum using two different approaches can be significant. Recent experimental observations at RHIC and the LHC indicate that charm quarks interact with the QGP very strongly so they lose significant amount of energy and also gain sizable collective flow. This could indicate that the condition for the Langevin approach for describing charm quark in-medium evolution may not be satisfied. To better constrain the uncertainty of the heavy quark diffusion coefficient, the next step would be to carry out precision measurement of bottom production in heavy-ion
collisions.
}

\item{{\bf Hadronization: fragmentation vs. coalescence?}

Coalescence hadronization in the strongly coupled QCD medium will lead to re-distribution how charm quarks form final state charm hadrons compared to the fragmentation hadronization anticipated in $p+p$ collisions. Recent experimental results show that coalescence hadronization can qualitatively explain the enhancement in charm baryon and charm-strangeness meson production observed in heavy-ion collisions. However, it remains still a challenge for many phenomenology models to quantitatively reproduce the experiment data of $\Lambda_c^+$ and $D_s^+$ production in heavy-ion collisions. In addition, the observed $\Lambda_c^+/D^0$ ratio in mid-central Au+Au collisions is quite larger than that from thermal model predictions, and the $\Lambda_c^+/D^0$ ratios in high energy $p+p$ and $p+$A collisions are already significantly larger than the fragmentation baseline. It will still need detail investigation on the charm baryon and charm-strangeness production in high energy nucleus nucleus collisions to better understand the hadronization scheme happening during the evolution.
}
\end{itemize}

With the unprecedented precision in the experimental measurements of charm hadron $R_{\rm AA}$ and $v_2$, it is then an important step forward to combine the experimental data with theoretical model calculations to extract the heavy quark spacial diffusion coefficient. However, there are considerable differences between different approaches while many of them can all reasonably describe the single charm hadron $R_{\rm AA}$ and $v_2$ data.  We ought to compare different phenomenological model approaches, separate out the trivial (e.g. initial condition, bulk medium evolution etc.) and non-trivial differences (e.g. perturbative vs. non-perturbative, Langevin vs. Boltzmann, different hadronization schemes etc.) between different models. Recently theory collaboration initiatives EMMI-RRTF~\cite{Rapp:2018qla} and JET-HQ~\cite{Cao:2018ews} have started to investigate and disentangle the differences between different models with aim to have a coherent understanding of the heavy quark in-medium diffusion coefficient.

The next phase of heavy quark program will be focusing on the following critical measurements: a) precision measurement of open bottom production particularly down to the low momentum region for detail investigation of mass dependence of parton energy loss as well as to precisely determine the heavy quark spacial diffusion coefficient parameter and its temperature dependence; b) precision measurement on heavy flavor baryon states for detail study of heavy quark hadronization and in-medium interactions; and c) heavy quark correlation measurement as a new sensitive and complementary probe to parton energy loss mechanism and medium transport parameter.

These measurements require significantly large datasets with better tracking detectors. The ALICE ITS-upgrade at the LHC and the sPHENIX MVTX detector utilizing the next generation fast MAPS sensor are planned to explore these measurements at the LHC and RHIC. The key feature of fast MAPS sensor upgrades compared to the STAR pixel MAPS detector is that the read-out integration time can be improved by a factor of 10--40. The technology advance will significantly reduce the background hits in high luminosity heavy-ion collisions and therefore greatly improve the tracking efficiency for low momentum tracks. These will enable precision measurements of physics observables outlined above that are needed to fulfill our physics mission.


\bibliographystyle{h-physrev}
\bibliography{HQinHIC}


\end{document}